\documentclass[11pt,fleqn]{article}
\usepackage{graphicx}
\usepackage{epsfig}
\usepackage{latexsym}

\newcommand{\freeFlight}{mean free flight time}

\newcommand{\rf}     [1] {~\cite{#1}}
\newcommand{\refref} [1] {ref.~\cite{#1}}
\newcommand{\refrefs}[1] {refs.~\cite{#1}}
\newcommand{\refeq}  [1] {(\ref{#1})}
\newcommand{\reffig} [1] {fig.~\ref{#1}}
\newcommand{\refFig} [1] {Fig.~\ref{#1}}
\newcommand{\reftab} [1] {table~\ref{#1}}
\newcommand{\refsect}[1] {sect.~\ref{#1}}
\newcommand{\refappe}[1] {appendix~\ref{#1}}

\newcommand{\beq}{\begin{equation}}
\newcommand{\continue}{\nonumber \\ }

\newcommand{\eeq}{\end{equation}}
\newcommand{\ee}[1] {\label{#1} \end{equation}}
\newcommand{\bea}{\begin{eqnarray}}

\newcommand{\eea}{\end{eqnarray}}

\newcommand{\combinatorial}[2]{ {#1 \choose #2}}

\newcommand{\FIG}[4]{\begin{figure}
                      \centering{#1}
                      \caption[#2]{#3}
                      \label{#4} \end{figure} }

\newcommand{\ie}{{that is}}             

\newcommand{\dzeta}{dyn\-am\-ic\-al zeta func\-tion}
\newcommand{\Fd}{spec\-tral det\-er\-min\-ant}

\newcommand{\obser}{a}          
\newcommand{\Obser}{A}          
\newcommand{\msr}{{\mu}}                        
\newcommand{\zetaInv}{{1/\zeta}}
\newcommand{\expct}    [1]{\left\langle {#1} \right\rangle}
\newcommand{\timeAver} [1]{\overline{#1}}
\newcommand{\eigenvL}{{s}}        

\newcommand\flow[2]{{f^{#1}(#2)}}
\newcommand\fullflow[2]{{{\bf f}^{#1}(#2)}}
\newcommand\xInit{{x_0}}                
\newcommand\pSpace{x}           
\newcommand\fullpSpace{{\bf x}}
\newcommand{\pS}{{\cal M}}                      
\newcommand{\PoincS}{{\cal P}}                  

\newcommand\period[1]{{T_{#1}}}                 
\newcommand{\cl}[1]{{n_{#1}}}   
\newcommand{\nCutoff}{N}        
\newcommand{\ktrunc}{M}

\newcommand\sumprime{\mathop{{\sum}'}}
\newcommand{\pseudos}{\pi}

\begin{document}

\title{Periodic orbit sum rules for billiards: \\ Accelerating cycle expansions}

\author{Sune F. Nielsen, Per Dahlqvist\\ 
        Royal Institute of Technology,
        S-100 44 Stockholm, Sweden \\
                   {\em and}\\
        Predrag~Cvitanovi{\a'{c}}\\
        Department of Physics \& Astronomy,
        Northwestern University\\
        2145 Sheridan Road,
        Evanston, IL 60208-3112
        }
\date{}
\maketitle

\\
{\bf keywords:} cycle expansions, periodic orbits, \dzeta s, 
sum rules, billiards.
 
\begin{abstract}
 We show that the periodic orbit sums for 2-dimensional billiards
 satisfy an infinity of exact sum rules. We test such sum rules and
 demonstrate that they can be used to accelerate the convergence 
 of cycle expansions for averages such as Lyapunov exponents. 
\end{abstract}

\section{Introduction}

Periodic orbit theory is a powerful tool for description of 
chaotic dynamical systems\rf{gaspard,gutbook,cvweb}.
However, as one is dealing with infinities of cycles, the formal theory
is not meaningful unless supplemented by a theory of convergence 
of cycle expansions.
For nice hyperbolic systems, the theory is well developed, and shows that
exponentially many cycles suffice to estimate chaotic averages
with super-exponential accuracy\rf{grothi,Rugh92,CRR93}.
However, for generic dynamical systems with infinitely specified 
grammars and/or non-hyperbolic phase space regions, the convergence of
the \dzeta s and \Fd s cycle expansions is less remarkable.
The infinite symbolic dynamics problem is generic,
and a variety of strategies for dealing with it have been proposed:
stability truncations\rf{DM97,DR},
approximate partitions\rf{hansen_thesis},
noise regularization \rf{noisy_Fred}
and even abandoning the periodic theory altogether\rf{cvbeyond}.

Computation of periodic orbits for a given system is often 
a considerable investment, as
locating exhaustively the periodic orbits of increasing length
for flows in higher dimensions can be a demanding chore.
It is therefore essential that the information 
obtained be used in an as effective way as possible. Here we propose a new, hybrid approach
of combining cycle expansions with exact results for ``nearby'' averages, based on 
the observation that the
periodic orbit sums sometimes satisfy exact sum rules. 

Studies of convergence of cycle expansions,
such as comparisons\rf{AACII} of truncation errors
of the dimension and the topological entropy for the H\'enon attractor,
indicate strong correlations in truncation 
errors for different averages. We propose to turn these
correlations in our favour, by using the error known exactly
by a sum rule to improve the estimate for a nearby average
for which no exact result exist. Billiards provide 
a convenient, physically motivated testing ground for this idea.
The approach is inspired by the formula
\refeq{MeanFree}
for \freeFlight\ in billiards, so well known to the Russian
school that it went unpublished for decades\rf{chernov}.
In this paper we show that billiards  
obey an infinity of exact periodic orbit sum rules,
and indicate how such rules might be used to accelerate
convergence of cycle expansions.

The paper is organised as follows: \refsect{cycle_avr} 
is a brief summary of the theory of periodic orbit averaging. 
In \refsect{sumrules} 
we review the known exact sum rules for billiards, and then
generalise them to an infinity of sum rules.
In \refsect{Numtest} we present the conventional cycle expansion
numerical results for our test system, the overlapping three-disk billiard.
This system is hyperbolic and does not suffer from the intermittency effects
that plague billiards such as the stadium and the Sinai billiards, 
but is still ``generic'' in the sense that its symbolic dynamics is
arbitrarily complicated. In \refsect{resum} and \refappe{app_resum} we develop a method which utilizes
the flow conservation sum rule to
accelerate the convergence of cycle expansions, and apply the
method to our test system.

\section{Periodic orbit averaging}
\label{cycle_avr}

We start with a summary of the basic formulas
of the periodic orbit theory -
for details the reader can consult \refrefs{gaspard,cvweb}.

A flow $\fullpSpace \rightarrow \fullflow{t}{\fullpSpace},  x \in \pS$,
is a continuous mapping ${\bf f}^t : \pS \rightarrow \pS$
of the phase space $\pS$ onto itself, 
parameterised by time $t$.
On a suitably defined Poincar\'e surface of section $\PoincS$,
the dynamics is reduced to a return map 
\beq
\pSpace \rightarrow \flow{n}{\pSpace} \hspace{1 cm} \pSpace \in \PoincS 
\,, 
\ee{map_def}
where $n$ is the ``topological time'', the number of times the
trajectory returns to the surface of section.

A \dzeta\rf{grothi} associated with the flow $\fullflow{t}{\pSpace}$ is
defined as the product over all prime cycles $p$ 
\beq
\zetaInv(z,\eigenvL,\beta)=\prod_p \left ( 1 -t_p \right )
\ , \qquad t_p = t_p(z,\eigenvL,\beta )=
\frac{1}{|\Lambda_p|}
e^{\beta \Obser_p - \eigenvL \period{p} }
z^{\cl{p}}
\,,
\ee{zeta_def}
where
$\period{p}$, $\cl{p}$ and $\Lambda_{p}$
are the period, topological length
and stability of prime cycle  $p$, $\Obser_{p}$
is the integrated observable $\obser(\pSpace)$
evaluated on a single traversal of cycle $p$
\beq
\Obser_p= \left \{ 
\begin{array}{ll}
\displaystyle \int_0^{\period{p} } \obser\left(\fullflow{\tau}{\pSpace_0}
\right)d\tau & \mbox{(flows)} \\
\displaystyle \sum_{k=0}^{\cl{p}-1} \obser\left(\flow{k}{\xInit}\right)  & \mbox{(maps)} \\
\end{array} \right . \, \pSpace_0 \in p  \,  ,
\ee{obs_flow}
where
$\eigenvL$ is a variable dual to the time $t$,
$z$ is dual to the discrete ``topological'' time $\cl{}$,
and $t_p(z,\eigenvL,\beta)$ is the weight of the cycle $p$.

Classical averages over chaotic systems are given by 
{\em cycle expansions} constructed from derivatives of \dzeta s.
By expanding the product \refeq{zeta_def}
a \dzeta\  can be represented as a {cycle expansion}
\bea
1/\zeta   &=&  1 - \sumprime_\pseudos t_\pseudos
\continue   
t_\pseudos  &=& t_\pseudos(z,\eigenvL,\beta) 
            =  (-1)^{k_\pseudos+1} t_{p_1} t_{p_2} \dots t_{p_k}
        \continue   
        &=& 
(-1)^{k_\pseudos+1}
{1 \over |\Lambda_\pseudos|}
e^{\beta \Obser_\pseudos- \eigenvL \period{\pseudos}}
z^{\cl{\pseudos}}
\,.
\label{equ_zeta_pow}
\eea
where the prime on the sum indicates that the sum
is restricted to pseudocycles, built from  
all distinct products of non-repeating
prime cycles weights. The pseudocycle topological length,
period, integrated observable, and stability are
\bea
\cl{\pseudos}&=&\cl{p_1}+ \ldots + \cl{p_{k}} 
\,,\qquad
\period{\pseudos}=\period{p_1}+ \ldots + \period{p_{k}} 
        \continue
\Obser_\pseudos&=&\Obser_{p_1}+ \ldots + \Obser_{p_{k}} 
\,,\qquad
\Lambda_\pseudos = \Lambda_{p_1}\Lambda_{p_2} \cdots \Lambda_{p_{k}} 
\,,
\label{equ_notation}
\eea
where $k=k_\pi$ is the number of involved prime cycles.
For economy of notation we shall usually omit the explicit dependence
of $1/\zeta$ and  $t_p$ on ($z$, $\eigenvL$, $\beta$)
whenever the dependence is clear from the context.

Truncation of the \dzeta\ with respect to the topological length 
$\cl{\pseudos} \leq \nCutoff$
will be indicated by a subscript
\beq
\zetaInv_{\nCutoff}(z,\eigenvL,\beta) =
1- \sumprime_{\cl{\pseudos} \leq \nCutoff }
t_\pseudos
\,.
\ee{equ_trnk_notation}

If the system bounded (such that no trajectories escape), the \dzeta\ \refeq{zeta_def} has a leanding zero at $\zetaInv(1,0,0) = 0$.
Expressing this condition in terms of the cycle expansion
\refeq{equ_zeta_pow} we find that any bound system satisfies the {\em
  flow conservation} sum rule \cite{cvweb}:
\beq
\zetaInv(1,0,0)=1-\sumprime_\pseudos (-1)^{k_\pseudos+1}{1 \over |\Lambda_{\pseudos}|}
=0
\,.
\ee{norm}

The cycle expansions for the phase space average of observable $\obser(x)$
are given by either the integral over the natural measure, or by the
cycle expansions
\begin{eqnarray}
\mbox{flows:  } \expct{\obser}_{\mbox{\footnotesize{flow}}}&=&\int_\pS \obser(\pSpace) \rho(\pSpace) d{\pSpace}=
\frac{\expct{A}_\zeta}{\expct{T}_\zeta}
\label{A_time_aver} \\
\mbox{maps:  }
\expct{\obser}_{\mbox{\footnotesize{map}}}
        &=&
        \int_\PoincS \obser(\pSpace) \rho(\pSpace) d\pSpace
        =
\frac{\expct{A}_\zeta}{\expct{n}_\zeta},
\label{A_n_aver} \\ \nonumber
\end{eqnarray}
where $\rho $ denotes the natural measure.
As we shall show in \refeq{MeanFree} below, the averages computed from the two representations of dynamics are
related by the \freeFlight.

The cycle expansions required for the evaluation of periodic orbit
averages \refeq{A_time_aver} and \refeq{A_n_aver} are given by 
derivatives of the \dzeta\ with respect to $\beta$, $\eigenvL$ and $z$ 
\bea
\expct{A}_\zeta &=& \frac{\partial}{\partial \beta}\zetaInv(1,0,0) =\sumprime_\pseudos (-1)^{k_\pseudos+1} 
\Obser_\pseudos/|\Lambda_\pseudos|   \label{beta_deriv} \\
\expct{T}_\zeta &=& -\frac{\partial}{\partial s}\zetaInv(1,0,0)= \sumprime_\pseudos (-1)^{k_\pseudos}
T_\pseudos/|\Lambda_\pseudos|   \label{s_deriv} \\
\expct{n}_\zeta &=&\frac{\partial}{\partial z}\zetaInv(1,0,0) = \sumprime_\pseudos (-1)^{k_\pseudos+1}
n_\pseudos/|\Lambda_\pseudos|
\,.  
\label{n_deriv} 
\eea

\section{Periodic obit sum rules for billiards}   
\label{sumrules}

We start by reviewing the \freeFlight\ sum rule for billiards discussed by
Chernov in \refref{chernov}.

In a $d$-dimensional billiard, a point particle moves freely inside a
domain $Q$, scattering elastically off its boundary $\partial Q$.
The billiard flow $f^t$ on $\pS=Q \times S^{d-1}$ (where $S$ is the
unit sphere of velocity vectors) has a natural Poincar\'e surface of
section associated with the boundary
\beq
\PoincS = 
\partial{\pS} =
\left\{ \left( q,v\right) \in \pS:q \in \partial Q \mbox{ and }
v\cdot n(q) \geq 0\right\}
\ee{poincare}
where $n(q)$ is the inward normal vector to the boundary at $q$, defined
everywhere except at the singular set $\partial{\pS}^{*}$ of
nondifferentiable points of the boundnary (corners,cusps,etc).
In what follows we shall restrict the discussion to (two-dimensional)
billiards.

Assume that the particle has unit mass and moves with unit velocity,
$p_1^2+p_2^2=1$. The cartesian coordinates and their conjugate momenta for the
full phase space $\pS$ of the billiard are
\[
\fullpSpace=\left(q_1,q_2,p_1,p_2\right)
   =\left(q_1,q_2,\sin\phi,\cos\phi\right)
\, .
\]
Let the Poincar\'{e} map be the boundary-boundary map
$f: \partial{\pS} \to \partial{\pS}$, and 
parametrise the  boundary
$\partial{\pS}$ by the Birkhoff (area preserving) coordinates
\[
\pSpace=\left(s,p_s \right), \qquad p_s=\sin \theta ,
\]
where $s$ is the arclength measured along the boundary,
$\theta$ is the scattering angle measured from the outgoing
normal, and $p_s$ is the component of the momentum parallel
to the boundary. Both the area of the billiard $A=|Q|$ and its
perimeter length $L=|\partial Q|$ are assumed finite. Let $\tau(x)$ be
the time of flight until the next collision.
The continuous trajectory is parametrised by the Birkhoff coordinates 
together with a time coordinate $0<t<\tau(x)$
measured along the ray emanating from the boundary point
$x=(s,p_s)$.

The period of a cycle $p$ is the sum of the finite free-flight
segments
\[
\period{p }=\sum_{k=0}^{\cl{p}-1} \tau \left (f^k(x_0) \right),
\]
where $x_0=(s_0,p_{s,0})$ is any of the collision points in cycle $p$.
The {\em \freeFlight } is the 
average time of flight between successive bounces off the billiard boundary.
It can be expressed either as a time average
\[
{\bar \tau}(x_0)=\lim_{n\rightarrow\infty}
{1\over n} \sum_{i=0}^{n-1} \tau(f^i(x_0)),
\]
or, as a phase space average
\beq
\expct{\tau }=\int_{\PoincS} \tau (x)d\msr(x),  
\ee{equ_m_f_p_def}
where $d\msr(x)=ds\;dp_s/\int_{\PoincS} ds\;dp_s$ is the natural
measure. For Hamiltonian flows like the billiard flow considered here
this is simply the Lebesgue measure. If the billiard is ergodic, the
time average is defined and independent of $x_0$ for almost all $x_0$.
In order to find an exact expression for the phase space average $\expct{\tau}$,
compute the integral over the entire phase space of the billiard,
\[
\int_\pS \delta(1-p_1^2-p_2^2) dq_1 d q_2 dp_1 dp_2=2\pi A
\]
and recompute the same thing in the 
Birkhoff coordinates,
\bea
\int \delta(1-p_1^2-p_2^2) dq_1 d q_2 dp_1 dp_2 &=&\int_{\PoincS} ds dp_s 
\int_{t=0}^{\tau(x)} dt
=\int_{\PoincS}  \tau(x) ds dp_s \continue
&=& \expct{\tau} \int_{\PoincS} ds dp_s=2 L \expct{\tau} , 
\eea
where $L$ is the circumference of the billiard. Hence the mean free
flight time is a purely geometric property of the billiard,
\beq
\expct{\tau}=\pi \frac{A}{L},
\ee{MeanFree}
the ratio of its perimeter to its area. 
The relation is a consequence of the Liouville measure being constant
and apply to any billiard regardless of whether its phase space is
mixed or not. 
For ergodic systems the
periodic orbit theory 
gives a cycle expansion formula \refeq{A_n_aver} for the
\freeFlight

\beq
\expct{\tau}=\frac{\expct{T}_\zeta}{\expct{n}_\zeta}.
\ee{MeanFreeCyc}
If we know $\expct{\tau}$ this formula enables us to
relate any discrete time average \refeq{A_n_aver}  computed from the
map to the continuous time averages \refeq{A_time_aver} computed on the flow.  
They are linked by the \freeFlight\ formula
\beq
\expct{a}_{map}=\expct{a}_{flow} \expct{\tau}.
\ee{mapflowrel}

In what follows we will restrict our attention to map
averages, and omit the subscript $\expct{\ldots}_{map}\rightarrow\expct{\ldots}$.

As the next example of a periodic orbit sum rule, consider the case of the observable being the
transverse momentum change at collision, $a=2\cos \theta$. 
The corresponding sum rule is called {\em the pressure sum rule} because
it is related to the pressure exerted by the particle
on the  billiard boundary.

The average {pressure} is given by the relation $P=F/ \left| \partial Q\right|$, where $F$ is the
time average of momentum change, \ie\ the force the particle exerts against the boundary.
The momentum change per bounce equals twice the transverse momentum at
the collision, so the average force per bounce is  
\beq
\expct{F}_{map}=\int_{\PoincS}2 p_\perp(x) d\msr(x)=\frac1{\int_\PoincS
  dsdp_s}\int_{\partial Q}\int_{-\pi /2}^{\pi /2}2\cos \theta\cos \theta d\theta ds
=\frac{\pi}{2}
\, .
\ee{dp_dQ}
Hence the pressure for a flow becomes 
\beq
\expct{P}_{flow}=\frac{\expct{F}_{map}}{L\expct{\tau}}=\frac1{2A}.
\ee{P_res}

The exact averages \refeq{MeanFree}, \refeq{P_res} apply to billiards of any shape, ergodic or not.
As both the mean free path and pressure can be calculated by means of
cycle expansions, these relations leads to exact billiard sum rules.

Now we note that as the Liouville invariant measure of the map
is constant,
{\em any average} of an observable $a(\pSpace)$, 
defined in terms of $\partial{\pS}$ coordinates 
$x=(s,p_s)$ can
be expressed in terms of a simple integral.
For each such observable we obtain an exact periodic orbit sum rule
\beq
\expct{a}=\frac{\expct{A}_\zeta }{\expct{n}_\zeta}
=\int_{\PoincS} a(x)d\mu (x).
\ee{analytalos}

Surprisingly enough, this uncountable infinity of sum rules 
seems not to have been noted in the literature.

The formula \refeq{analytalos} does not allow for analytical
computation of every average we want to compute in a billiard. 
Consider the simplest nontrivial average worth study in billiards, the
diffusion coefficient
\beq
\expct{D}=\frac1{2d}\frac1{\expct{n}_\zeta}\frac{\partial^2}{\partial\beta^2}
\left. (\zetaInv) \right |_{\beta=0}.
\ee{diff_def}
It requires evaluation of a second derivative of the relevant 
\dzeta, this means that two-point correlations of the
observable along cycles will enter the averaging formulas, so the average can not be computed from one iterate
of the map. 

Another quantity of interest is the Lyapunov exponent. Let $\Lambda(x_0,n)$ is the largest eigenvalue of the 
Jacobian of the $n$th iterate of the map. The (largest) Lyapunov
exponent is 
defined as 
\[
\lambda = \lim_{n \rightarrow \infty} \frac{1}{n} \log |
\Lambda(x_0,n)|\, .
\]

\noindent
The cycle expansion formulas in \refsect{cycle_avr} compute 
\beq
\lambda = \lim_{n \rightarrow \infty}  \frac{1}{n} 
\int d\rho(x) \log | \Lambda(x,n)|  \, ,
\ee{lyapunov_aver}
that is, a combination of time and phase space averages.
Note that if $\Lambda(x_0,n)$ is multiplicative,
$\Lambda(x_0,n)=\prod_{k=0}^{n-1}\Lambda(f^k(x_0))$, then the integral
in \refeq{lyapunov_aver} is independent of $n$: setting $n=1$, reduces
the problem to one iterate of the map.
This is the case for 1-dimensional maps. However in most cases the
invariant measure $\rho(x)$ is not known a priori, and there is no
simple exact formula for the average. 

For billiards the invariant density is known but 
the expanding stability eigenvalue $\Lambda(x_0,n)$ is
not multiplicative along an arbitrary trajectory,
, and the integral in \refeq{lyapunov_aver} is dependent on $n$.
It is possible to derive a multiplicative evolution operator for this
purpose \rf{pcgv}.
However, for the purpose at hand the naive cycle expansion formulas still apply, because 
$\Lambda(x_0,n)$ is multiplicative for repeats of periodic orbits.
By defining the cycle weight 
\[
e^{\beta \Obser_p}=|\Lambda_p|^\beta
\]
the cycle expansion for the Lyapunov exponent is given by
\beq
\expct{\lambda}=\frac{\expct{\ln{|\Lambda|}}_\zeta}{\expct{n}_\zeta}
\ee{Lyapunov_cyc}

So even though Lyapunov only
requires computation of two first order derivatives of the \dzeta, it
requires $n$-point correlations to all orders and cannot be computed by a sum rule. 

In the case when \refeq{mapflowrel} the Lyapunov exponent of the flow to
the Lyapunov exponent of the corresponding Poincar\'{e} return  map, 
the relation is known as the Abramov's formula \rf{Abramov}.

\section{Three overlapping disks billiard}
\label{Numtest}

\FIG{\includegraphics[width=0.50\textwidth]{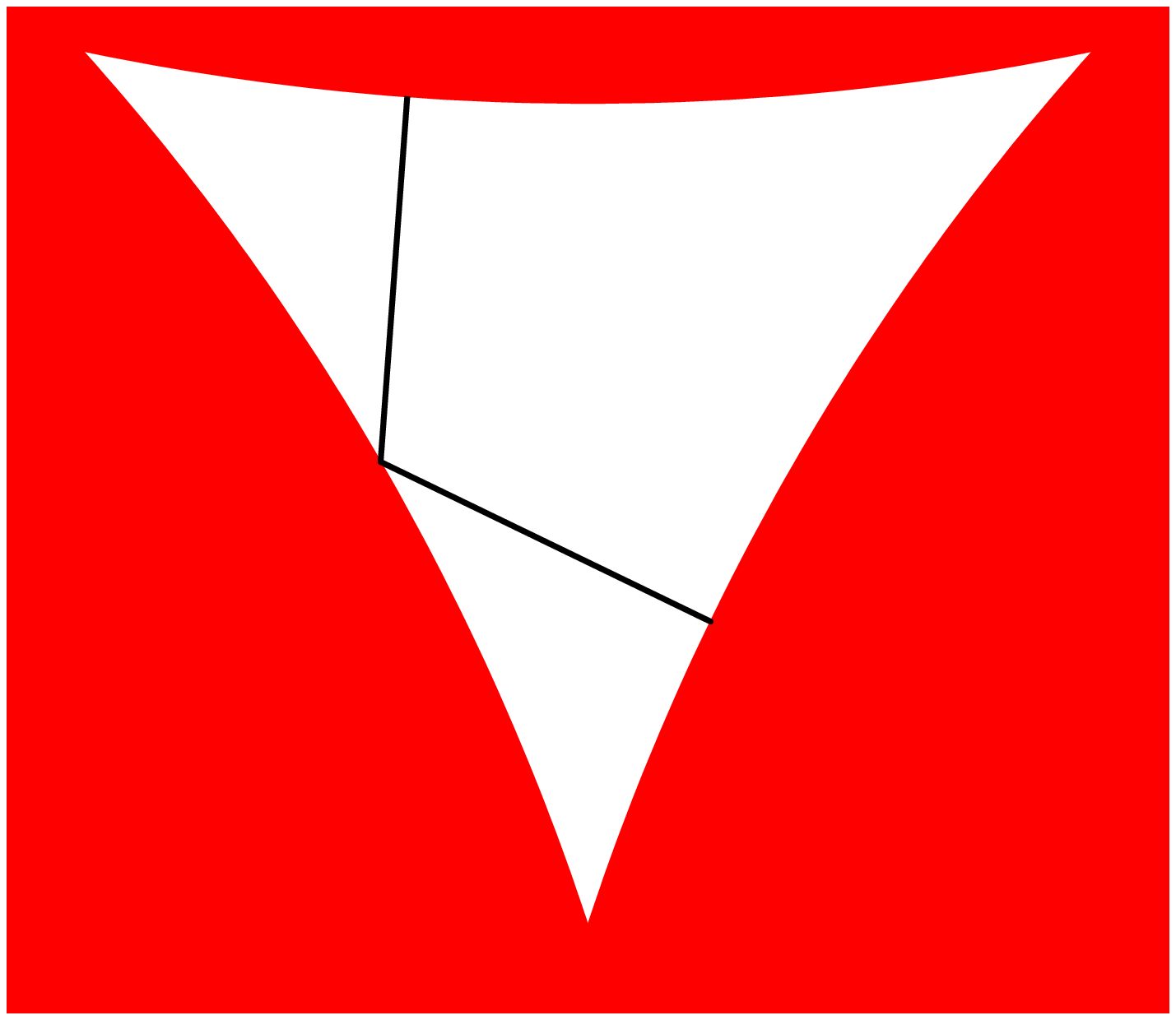}
}{}{
The overlapping 3-disk billiard. A point-like particle moves 
inside the billiard bouncing specularly off the boundary. Shown is a
cycle of topological length 4.
}{fig_3disk_def}

We will test the above sum rules on cycle expansions for 
a concrete system, the
 overlapping 3-disk billiard. 
This billiard consists of three disks of radius $a$ centered on the corners
of an equilateral triangle with sides $R$.
There is a finite enclosure (see \reffig{fig_3disk_def}) between the disks
if $\sqrt{3}<R/a \leq 2$. This enclosure defines the billiard domain
$Q$. One of the limits
$R/a \rightarrow \sqrt{3}$ 
corresponds to the integrable equilateral triangle billiard.
The other limiting 
case
$R/a=2$ exhibits intermittency with infinite sequences of periodic
orbits whose periods $T_p$ accumulate to finite limits, and where stabilities
 fall off as some power $n_p^\alpha$, where $n_p$ is the topological length.

The $C_{3v}$ symmetry of the billiard enables us to work in the fundamental domain \cite{CE93}. The fundamental
domain is one 6th slice of the billiard domain, fenced in by the symmetry axes of the billiard.
In what follows we are only interested in the lowest eigenvalue and
therefore we restrict our computations to the fully symmetric $A_1$ subspace.
The fundamental domain symbolic dynamics is binary, but is not of the
finite subshift type; its full specification would require an infinity
of pruning rules of arbitrary length.

The \freeFlight\ \refeq{MeanFree} for the 
overlapping 3-disk billiard can be found by geometric considerations,
\beq
\expct{\tau} =
        \pi \frac{R^2/4\sqrt{3}-a^2\theta-Rr/2}
        {2a \theta},
\ee{exact_mfp}
where
$r=\sqrt{a^2-(R/2)^2}$ and $\theta=\pi/6-\arcsin(r/a)$.
We shall set $a=1$ throughout this paper, and parameterise the
billiard by the center-to-center distance $R$. 
All our numerical tests are done for $R = 1.9$. 
Results for this parameter value, as well as for $R=1.85$ are shown in \reftab{t:BestEst}.

\begin{table}
\begin{tabular}{|l|llll|r|}
\hline
~$R$  & ~~$\expct{\tau}$& ~~$\expct{\lambda}$& $
  \lambda_{\mbox{num}}$ & ~~$L \expct{P} $  & \# cycles
\cr \hline
    1.85  & 0.102  &  0.523 & &  1.57 &   342
\cr 1.9   & 0.1401 &  0.6036 & 0.60363 &  1.570  &   525 
\cr \hline
\end{tabular}
\caption[]{\small
The \freeFlight\ $\expct{\tau}$ , the average
pressure $P$ , and the best estimate
of the Lyapunov exponent $\lambda$ computed by cycle expansion as function of 
the 3-disk center-to-center separation $R$ used in our numerical
tests, with disk radius fixed to $a=1$. For $R=1.9$ a numerically
computed Lyapunov reference value obtained by direct simulation using $10^{10}$
bounces is displayed.
The total numbers of the fundamental domain prime cycles used 
in the cycle expansion computations are also indicated.
}
\label{t:BestEst}
\end{table}

\refFig{fig_sumr_convg} illustrates the convergence of finite
topological length cycle expansions for the flow
conservation sum rule \refeq{norm} and for the \freeFlight\ sum rule 
\refeq{MeanFreeCyc}.
As the exact result is known, we plot the logarithm of the error
as function of the truncation length N. 

\FIG{\includegraphics[width=0.40\textwidth]{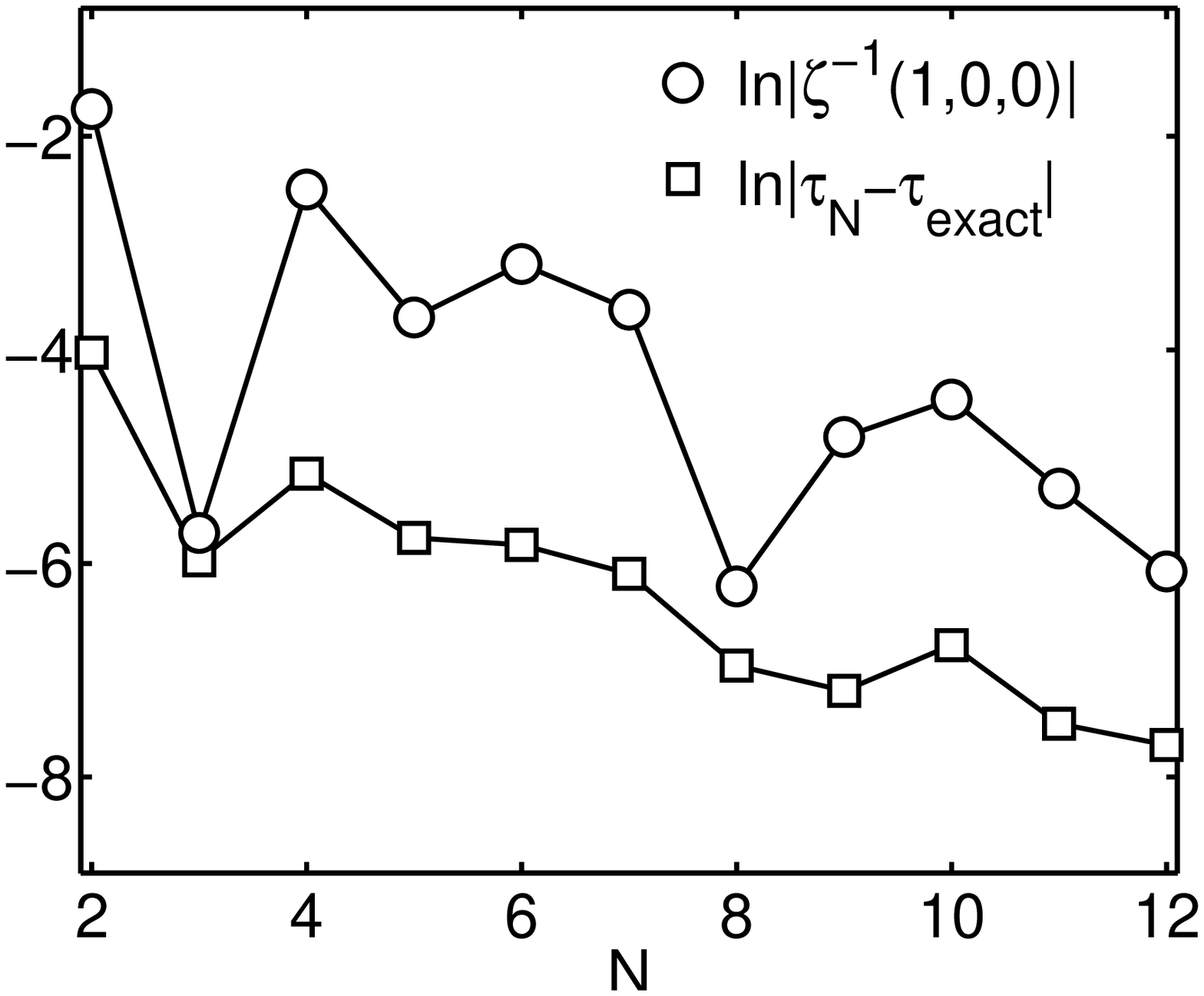}
}{}{
Convergence of cycle expansions:  
Deviation of  cycle expansions truncated to the topological length $N$
from exact sum rules
for ($\circ$) flow conservation \refeq{norm} and 
($\Box$) \freeFlight\ \refeq{MeanFreeCyc}. 
}{fig_sumr_convg}

The overall exponential convergence reflects the existence of a gap,
the \dzeta s are analytic beyond $z=1$.
The ``irregular'' oscillations in \reffig{fig_sumr_convg} are typical
for systems with complicated symbolic dynamics. For systems with
finite subshift symbolic dynamics the oscillations ceases when
the cutoff $\nCutoff$ exceeds the longest forbidden substring and
if the full \Fd\ is used, super-exponential convergence
sets in \cite{AACII}.

One should note the coincidence of the peaks and dips of the two
curves. This type of correlations between coefficients between different
power series will be important in the following.

\section{Utilizing exact sum rules}
\label{resum}

We shall illustrate the utility of exact sum rules in accelerating the 
convergence of cycle expansions by applying flow conservation sum rule
to the problem of computing the {\freeFlight}\ \refeq{MeanFree}.
As we already have the exact formula for this average, we will be able to compute the exact error in the various estimates and compare them.
We will then apply the same technique to evaluate the Lyapunov exponent, for which no exact formula exists.

The idea is to use the flow conservation sum rule \refeq{norm} to improve the numerator and denominator
of \refeq{MeanFreeCyc}  separately.

We begin by the denominator.
The general problem is to 
find a good
estimate of the derivative $F'(z_0)$ at the first root $F(z_0)=0$ of a function given by a power series $F(z)=\sum_{k=0}^\infty b_k z^k$
where the coefficients are known only up to order $N$
\begin{equation}
F_{(N)}(z)=\sum_{k=0}^N b_k z^k ,
\end{equation}
and we know (\refappe{app_resum}) that a good estimate is given
by
\begin{equation}
F'(z_0) \approx F'_{(N)}(z_0)-\frac{N+1}{z_0}  F_{(N)}(z_0) . \label{equ_fp_resum}
\end{equation}
We argue in  \refappe{app_resum} that the error in the
above estimate is suppressed  compared to the error of
the estimate $F'_{(N)}(z_0)$ by a factor $q$
whose asymptotic behaviour is
\begin{equation}
q \sim 1/N.
\end{equation}

The improved estimate of the derivative of the zeta function \refeq{n_deriv}
thus reads
\begin{equation}
\expct{n}_{\zeta,acc} \approx 
\expct{n}_{\zeta,(\nCutoff)}-(\nCutoff +1) 
\zeta^{-1}_{(\nCutoff )}(1,0,0).
\end{equation}

The continuous time average by invoking the sum rule \refeq{norm} is similar to the previous one but $\zetaInv(1,\eigenvL,0)=0$
is now a Dirichlet series in $\eigenvL$.
The basic idea is to start by expanding the zeta function around some point
$\eigenvL=\eigenvL_0$
\[
\zetaInv(1,\eigenvL,0)
=1-\sumprime_{\pseudos} \frac{(-1)^{k_\pseudos-1}}{|\Lambda_\pseudos|} e^{-\eigenvL T_\pseudos} 
\]
\[
=1-\sumprime \frac{(-1)^{k_\pseudos-1}}{|\Lambda_\pseudos|} e^{-s_0 T_\pseudos} 
\sum_{k=0}^\infty \frac{(s_0-s)^k T_\pseudos^k}{k!}. 
\]
Since we have only a finite number of of pseudocycles  at our
disposal, it is not meaningful to consider 
high powers $k$ in the above Taylor series. 
So we truncate the series
\bea
\zetaInv_{(\nCutoff,\ktrunc)}(1,\eigenvL,0)&=&1-
\sumprime_{\pseudos: n_\pseudos \leq \nCutoff} 
\frac{(-1)^{k_\pseudos-1}}{|\Lambda_\pseudos|} e^{-s_0 T_\pseudos}  
\sum_{k=0}^M \frac{(s_0-s)^k T_\pseudos^k}{k!} \continue &=&
1-\sum_{k=0}^{M}\frac{\expct{T^k}_{\zeta,(N)}}{k!} (s_0-s)^k
\,. 
\label{equ_zeta_flow}
\eea
where
\[
\expct{T^k}_{\zeta,(N)}=\sumprime_{\pseudos: n_\pseudos \leq \nCutoff} (-1)^{k_\pseudos-1}
\frac{e^{-s_0 T_\pseudos}T_\pseudos^k}{|\Lambda_\pseudos|}
\]
is the $k$th moment of the average cycle time.
The sets of periodic orbits going into the calculation are still being truncated
according to their topological length.
This finite Taylor series is the analogue of the truncated function 
$F_{(N)}(z)$ treated above;
$s=0$ corresponds to $z=1$, and $s=s_0$ to $z=0$.
We can use \refeq{equ_fp_resum} and write down the improved estimate 
\beq
\expct{T}_{\zeta,acc} \approx 
\expct{T}_{\zeta,(\nCutoff)}-\frac{\ktrunc+1}{s_0} \zeta^{-1}_{(\nCutoff,\ktrunc)}(1,0,0)
\ee{res_flow}

However there is now an additional complication due to the fact that not all
available $\expct{T^k}_\zeta$ are exact. 
So how should we choose $s_0$ and $\ktrunc$? We choose $s_0$ to lie somewhere in the range
$1/T_{max}<s_0 < 1/T_{min}$ where the $T_{min}$ and $T_{max}$
are the smallest and largest period in the sample for a particular
topological length cutoff $\nCutoff$.
 
The next question is,
for a given $s_0$, what is the number of accurate coefficients 
$\expct{T^k}_{\zeta,(N)}/k!$?
We see from \refeq{equ_zeta_flow} that pseudocycles
are suppressed with their length
according to the function $T^k \exp(-s_0 T)$ having its maximum at
$T=k/s_0$. So the coefficients with $k\ll s_0 T_{max}$ can be expected
to be accurate. However, as
the majority of cycles have periods close to $T_{max}$ we want
to make use of the information they carry. We have found it preferable to include a large
number of fairly accurate coefficients rather
than a small number of very accurate ones.
So we choose the maximum power $\ktrunc$ to be given by
the average cycle length
\[
\ktrunc=s_0 \left.{\timeAver{T}}_p \right |_{\cl{p}=\nCutoff-1 }
\label{Mtruncation}
\]

\FIG{\includegraphics[width=0.40\textwidth]{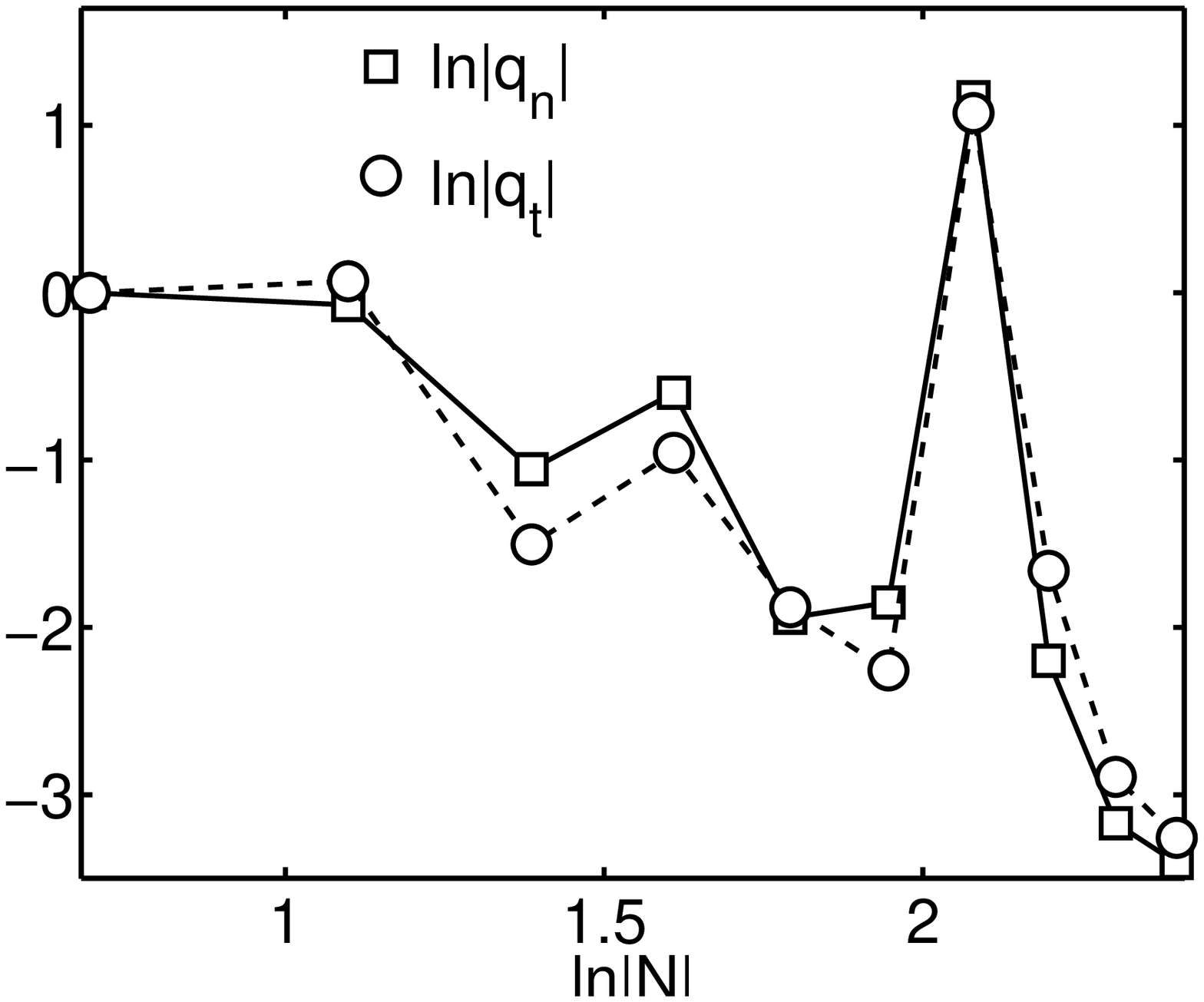}
}{}{
The error suppression factor for 
($\Box$) maps and for ($\circ$) flows, applied to the 3-disk system
with $R=1.9$. Here we have used an extrapolated value from the cycle
expansion as the best asymptotic estimate. Both error suppressions display the
estimated error decrease and demonstrate that the sum rules 
do improve convergence.
}{fig_qnqt}

The error of the improved estimate is suppressed
compared to the error of the traditional estimate
by factor we call $q$, see \refappe{app_resum}.
This $q$-factor
is plotted  in \reffig{fig_qnqt}. It decreases 
(apart from oscillations) as the
estimated $\nCutoff^{-1}$ error suppression derived for maps.

The calculation of the integrated observable amounts to evaluating the
$\beta$ derivatives of the \dzeta s. The role of $\beta$ is completely analogous to that of $s$.
With $\beta$ viewed as a complex variable, 
the \dzeta\ $\zetaInv(1,0,\beta)$ is a Dirichlet series in $\beta$ and
the above methods can be used to compute $\frac{\partial}{\partial
  \beta} \zetaInv(1,0,0)$.
Here similar criteria apply to $\beta_0$ and $N$ as for
\refeq{res_flow}
: $\beta_0$ close to $1/A_{min}$ and 
$M=\beta_0 \left.{\timeAver{A}}_p \right |_{\cl{p}=\nCutoff-1 }$.

\subsection{Improvement on the averages}
\label{results}

So far we have improved the numerator and the denominator 
of \refeq{MeanFreeCyc} and \refeq{A_n_aver}
separately. The errors of both are suppressed by a factor $q \approx O(1/N)$
compared to unaccelerated estimates.
We have also seen (\reffig{fig_qnqt}) that, 
both before and after resummation, their behavior versus the cutoff
$\nCutoff$ are
highly correlated.
So it is not obvious how the resulting average should be improved, indeed
it is not clear whether it is improved at all.

\noindent
The accelerated cycle expansion for an observable $\obser (x)$ using
our method is
\beq
\expct{\obser}_{acc}=\frac{\expct{A}_{\zeta,acc}}{\expct{n}_{\zeta,acc}},
\ee{equ_our_acc}
and the error suppression, the $q$-factor for the observable $\obser
(x)$ is 
\beq
q_a=\frac{\expct{\obser}_{acc}-\expct{a}_{exact}}{\expct{\obser}-\expct{a}_{exact}}.
\ee{equ_qave}

\FIG{\includegraphics[width=0.40\textwidth]{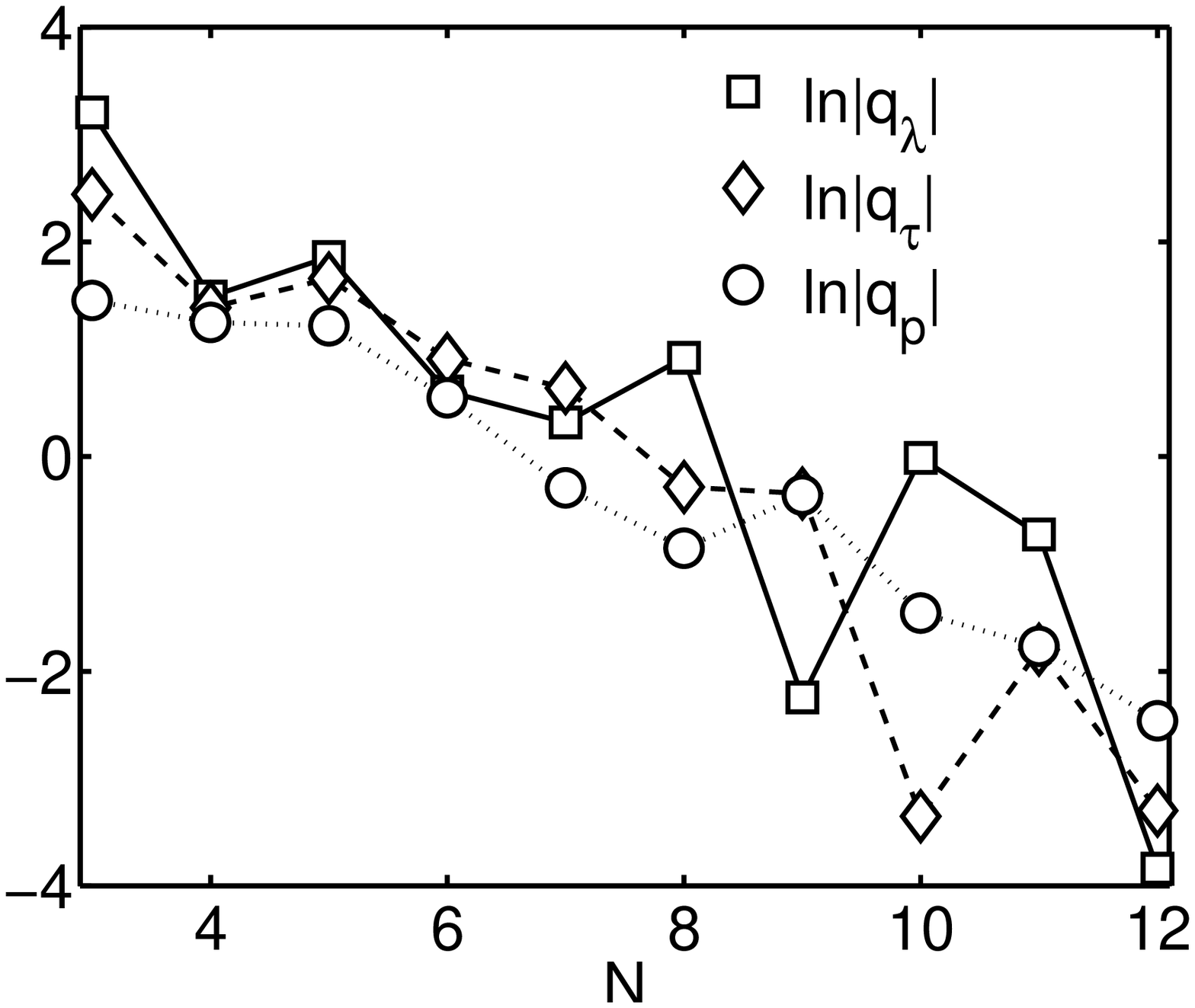}
}{}{
The error suppression factor \refeq{equ_qave} for : ($\Diamond$) The accelerated
 \freeFlight sum rule. ($\circ$)  pressure sum rule. 
($\Box$) The Lyapunov exponent. 
At topological length 12 the accuracy of the accelerated Lyapunov
exponent has reached the best estimate from direct numerical
simulation (see Table 1).
}{fig_conv}

We compute this $q$ factor numerically for three different averages: 
\begin{itemize}
\item[(i)] The \freeFlight\ $\expct{\tau}$ by \refeq{MeanFreeCyc}.
The exact result is given by  \refeq{MeanFree}.

\item[(ii)] The average force $\expct{F}_{map}$ by \refeq{dp_dQ},
associated
with the pressure sum rule. 
The exact result is given by  \refeq{P_res}. 

\item[(iii)] The Lyapunov exponent by \refeq{Lyapunov_cyc}.
The reference value of the Lyapunov exponent is obtained by numerical 
simulation, see Table 1.
\end{itemize}

The results are summarized in \reffig{fig_conv}. The accelerated cycle expansions are clearly better
than the standard cycle expansions.
The error suppression factors appear to
decrease exponentially, and therefore the acceleration
techniques has for the 3-disk system increased the correlation between
the expansions leading to a faster convergence for the averages.

\section{Conclusion}
\label{conclusion}

In this paper we have achieved two objectives: 
(i) We have derived an infinite number of exact periodic orbit sum 
rules for billiards \refeq{analytalos}. Such sum rules enable us to 
make exact computations of some statistical averages for billiards, 
such as the \freeFlight\ \refeq{MeanFree} and pressure \refeq{P_res}. 
(ii) We have derived the improved estimate \refeq{res_flow} which 
combines the flow conservation sum rule \refeq{norm} with the cycle
expansions. In order to measure the convergence acceleration, we have 
introduced the error suppression factor \refeq{equ_qave} that gauges 
the improvement of the accelerated cycle expansions relative to the 
unaccelerated ones. We thus demonstrate that exact sum rules can be 
used to accelerate convergence for observables for which no exact 
results exist, see \reffig{fig_conv}.

A challenge for the future is to utilize such infinities of sum rules
for billiards in the classical applications (other than the
Lyapunov exponent studied here), as well as in the semi-classical
applications of periodic orbit theory.

This work was supported by the Swedish Natural Science
Research Council (NFR) under contract no. F-AA/FU 06420-312
and no. F-AA/FU 06420-313. PD thanks NORDITA for partial support.
SFN thanks PD and KTH for hospitality.

\appendix
\section{Resummation of power series}
\label{app_resum}

Consider a function  
$F(z)$ 
given by a power series,
where only a finite number of coefficients are known.
\beq
F_{(N)}(z)=\sum_{k=0}^N b_k z^k .
\ee{equ_Nf_def}

We assume that
$F(z_0)=0$ for some $z_0$ and we wish to estimate the
first derivative  $F'(z_0)$ 
(and possibly higher derivatives)
as accurately as possible. 
The general problem is to transform the Taylor series around $z=0$ into
a Taylor series around $z=z_0$, and extract the desired coeficients.
This is done by the ansatz
\beq
\sum_{k=0}^{N}b_k z^k = \sum_{i=1}^{N+1}a_i (z-z_0)^i+O(z^{N+1}) 
\ee{equ_ansatz}

Note that the sum rule is built into this ansatz by
setting $a_0= 0$. Our aim is to determine $F'(z_0)=a_1$.
We keep the number of known and unknown coefficients equal so that the system of equations is solvable.

Expanding the right hand side \refeq{equ_ansatz} binomially
\begin{equation}
\sum_{i=0}^{N}b_i z^i=\sum_{i=1}^{N+1}a_i \sum_{j=0}^{i} z^j (-z_0)^{i-j} 
\combinatorial{i}{j}+O(z^{N+1})
\nonumber
\end{equation}
we obtain the linear system of equations
\begin{equation}
b_j=\sum_{\max (j,1)}^{N+1}
\combinatorial{i}{j} (-1)^{i-j} a_i \label{equ_lin_sys} , \qquad 1\leq
i \leq N+1 \mbox{ and } 0\leq j \leq N .
\end{equation}
In order to transform this to matrix equations where all indices range
from $1$ to $n\equiv N+1$ we define vectors
\begin{equation}
({\bf b})_i=z_0^{i-1} b_{i-1}, \qquad
({\bf a})_i=z_0^{i} a_{i}  \  \  ,
\end{equation}
and rewrite \refeq{equ_lin_sys} as 
\begin{equation}
{\bf b}={\bf M}{\bf a}, \qquad
({\bf M})_{ij}=\combinatorial{j}{i-1} , \qquad 1\leq i,j \leq n
(-1)^{j-i+1}.
\end{equation}
We use a convention that 
$\combinatorial{n}{m}=0$ if $m$ is out of range.
This system may readily be solved.
Define the matrix ${\bf L}$ by
\begin{equation}
({\bf L})_{ij}=\left\{ \begin{array}{cc}1 & i \geq j\\
                               0 & i <   j  \end{array}\right.
\end{equation}
Then
\begin{equation}
({\bf LM})_{ij}=(-1)^{i+j+1}\combinatorial{j-1}{i-1} \rightarrow
({\bf LM})^{-1}_{ij}=-\combinatorial{j-1}{i-1},
\nonumber
\end{equation}
and the explicit solution is
\begin{equation}
{\bf a}=({\bf LM})^{-1}{\bf Lb}
\end{equation}

In particular
\bea
({\bf a})_1 &=&- n\; ({\bf b})_1-(n-1)\; ({\bf b})_2- \ldots -1\; 
({\bf b})_n 
=-\sum_{k=1}^{n}(n-k+1)({\bf b})_k \continue
z_0 a_1&=&({\bf a})_1=-\sum_{k=0}^{N}(n-k)z_0^k b_k=z_0 F'_{(N)}(z_0)-(N+1)F_{(N)}(z_0)
\eea
So our improved estimate of $F'(z_0)$ is
\begin{equation}
F'(z_0)\approx a_1=F'_{(N)}(z_0)-(N+1)z_0^{-1}F_{(N)}(z_0)
\end{equation}

The error is suppressed by a factor
\begin{equation}
q=\frac{F'_N(z_0)-nz_0^{-1}\; F_N(z_0)-F'(z_0)}{F'_N(z_0)-F'(z_0)}=
\frac{\sum_{k=n}^{\infty}(k-n) b_k}
{\sum_{k=n}^{\infty}k b_k}
\end{equation}
To get this on a more handy form we use {\em summation by parts},
that is, we define
\begin{equation}
1/q=1+\frac{ns_n}{\sum_{k=n+1}^{\infty}s_k},\qquad s_k=\sum_{j=k}^{\infty}b_j
\end{equation}

If $F(z)$ is the \Fd\ for a $d$-dimensional Axiom-$A$ map
the coefficients of the power series expansion are super-exponentially
bounded 
\begin{equation}
C_a \Lambda^{-n^{1+1/d}}_a < |b_n| <C_b \Lambda^{-n^{1+1/d}}_b
\end{equation}
where $1 < \Lambda_b < \Lambda_a$.
Assuming moreover that the signs of the coefficients settle down to some
periodic pattern, one can show that the error supression factor has the following asymptotic behaviour
\begin{equation}
q\sim n^{-(1+1/d)}
\end{equation}

In this paper we focus on a hyperbolic systems whose symbolic dynamics
cannot be finitely specified. In that case the bound on the 
coefficients is exponential
\begin{equation}
C_a \Lambda^{-n}_a < |b_n| <C_b \Lambda^{-n}_b,
\end{equation}
and nothing can be said about the signs, as they can oscillate in a completely
irregular fashion\rf{PDprun}.
It seems difficult to obtain proper bounds on $q$ in a general
setting. In the case at hand we can only provide a qualified guess on the decrease of
the error supression factor
\begin{equation}
q\sim n^{-1}
\end{equation}

\FIG{\includegraphics[width=0.40\textwidth]{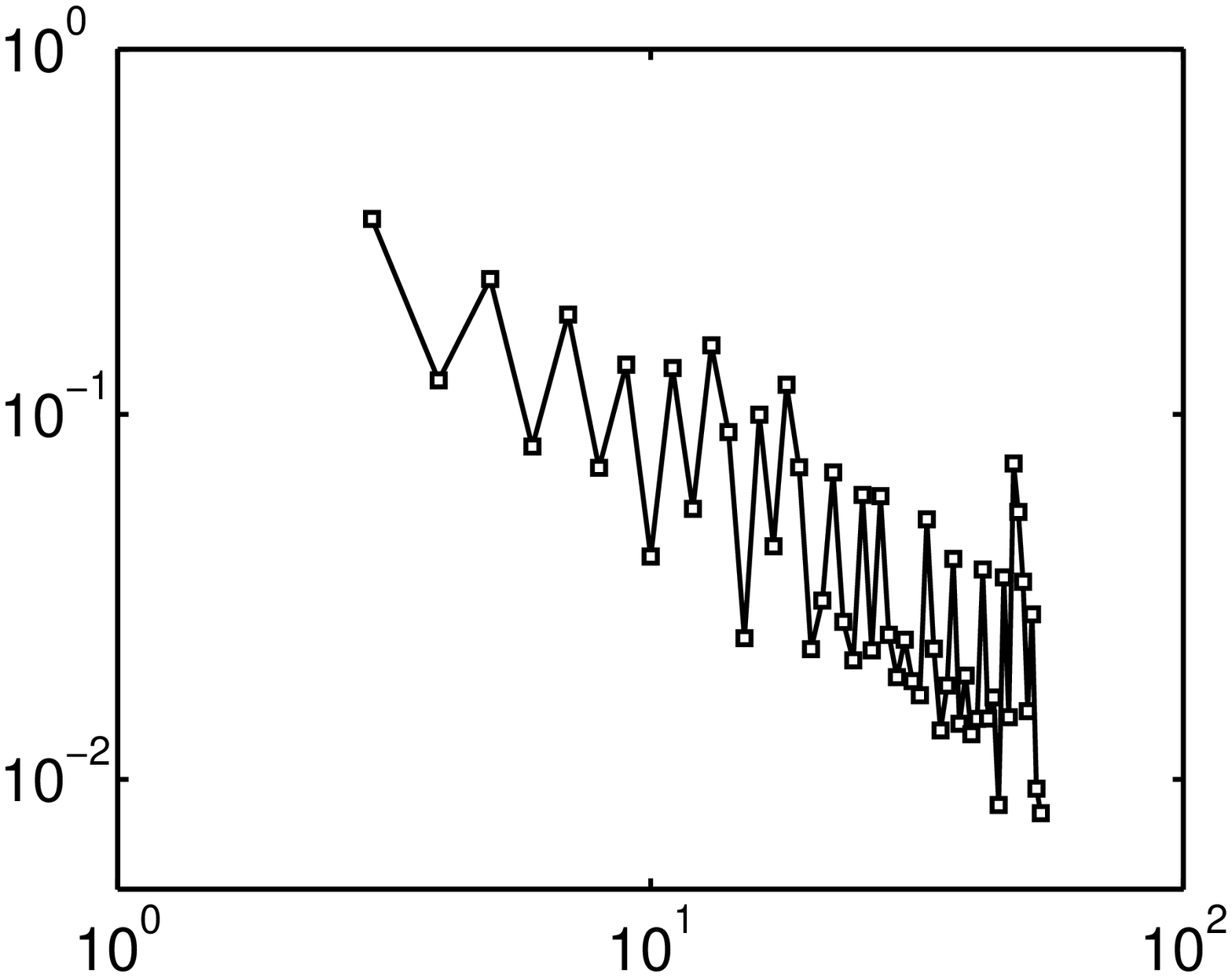}
}{}{
Error suppression factor \refeq{equ_qave} (y-axis) versus truncation in topological length
\nCutoff (x-axis) for the tent map (with a ``typical'' slope value $\Lambda$).
}{fig_tent}

Some evidence for this behavior can be provided by the tent map
\beq
F(x)=\left \{ 
\begin{array}{c}
\Lambda x  \mbox{ for }  x<\frac12 \\
-\Lambda(x-1) \mbox{ for } x\geq\frac12  \\
\end{array}\right . .
\ee{tent_map}
The expansion rate is uniform but complete symbolic dynamics is lacking
in the generic case. In \reffig{fig_tent} 
we plot the q-factor for the tent map
for a randomly chosen parameter versus $N$. It conforms with
the predicted $1/N$ behavior.

The ansatz \refeq{equ_ansatz}
used here is the simplest conceivable and it led to
very simple formulas. The only requirement is that the 
\dzeta\ is analytic
in a disk $z\leq R$ where $R>1$. This excludes strongly intermittent
systems where a more refined ansatz is needed\cite{PDresum}.
If one has some explicit knowledge of the nature of the leading
singularity of the \dzeta, one can taylor a more specific ansatz.



\end{document}